\documentclass[twocolumn,lineno]{aastex63}

\usepackage{amssymb,amsmath,verbatim,mathtools,needspace,enumitem,etoolbox,graphicx,physics,microtype,afterpage,tabularx,color,textcmds}
\usepackage{multirow}

\definecolor{linkcolor}{rgb}{0.0,0.3,0.5}
\usepackage[all]{hypcap}
\usepackage[T1]{fontenc}
\usepackage[utf8]{inputenc}
\usepackage{}

\def\apjl{\rm{ApJ}}
\def\apjs{\rm{ApJS}}
\def\aap{\rm{A\&A}}

\newcommand{\hubble} {Hubble-Lema\^{i}tre\,}
\newcommand{\plusNetThree} {HLV+}
\newcommand{\plusNetThreeS} {HLV+ }
\newcommand{\plusNet} {HLVKI+}
\newcommand{\plusNetS} {HLVKI+ }
\newcommand{\voyNet} {Voy+}

\newcommand{\initNet} {ECC}

\newcommand{\gwOhEight} {GW190814}
\newcommand{\gwOhEightS} {GW190814 }
\newcommand{\gwOhNine} {GW150914}

\newcommand{\gwtc} {GWTC-1}
\newcommand{\gwtcS} {GWTC-1 }

\begin{document}

\title{Dark Sirens to Resolve the Hubble-Lema\^{i}tre Tension}

\shorttitle{Dark Sirens to Resolve the Hubble-Lema\^{i}tre Tension}
\shortauthors{Borhanian et al.}

\newcommand{\penncosmos}{\affiliation{Institute for Gravitation and the Cosmos, Department of Physics, Pennsylvania State University, University Park, PA, 16802, USA}}
\newcommand{\miss}{\affiliation{Department of Physics and Astronomy, The University of Mississippi, Oxford, MS 38677, USA}}
\newcommand{\chennai}{\affiliation{Chennai Mathematical Institute, Siruseri, 603103, India}}
\newcommand{\pennastro}{\affiliation{Department of Astronomy \& Astrophysics, Pennsylvania State University, University Park, PA, 16802, USA}}
\newcommand{\cardiff}{\affiliation{School of Physics and Astronomy, Cardiff University, Cardiff, UK, CF24 3AA}
}

\correspondingauthor{Ssohrab Borhanian}
\email{sub284@psu.edu}

\author[0000-0003-0161-6109]{Ssohrab Borhanian}
\penncosmos 

\author[0000-0001-9930-9101]{Arnab Dhani}
\penncosmos 

\author[0000-0002-5441-9013]{Anuradha Gupta}
\miss

\author[0000-0002-6960-8538]{K. G. Arun}
\chennai

\author[0000-0003-3845-7586]{B. S. Sathyaprakash}
\penncosmos \pennastro \cardiff

\begin{abstract}
The planned sensitivity upgrades to the LIGO and Virgo
facilities could uniquely identify host galaxies of \emph{dark
sirens}---compact binary coalescences without any electromagnetic
counterparts---within a redshift of $z=0.1$. This is aided by the higher order
spherical harmonic modes present in the gravitational-wave signal, which also
improve distance estimation. In conjunction, sensitivity upgrades and higher
modes will facilitate an accurate, independent measurement of the host galaxy's
redshift in addition to the luminosity distance from the gravitational wave
observation to infer the \hubble constant $H_0$ to better than a few percent in
five years. A possible Voyager upgrade or third generation facilities would
further solidify the role of dark sirens for precision cosmology in the future.
\end{abstract}

\keywords{Gravitational waves, dark sirens, cosmology, \hubble tension.}

\section{Introduction}

The \hubble constant $H_0$ is a fundamental cosmological quantity that governs
the expansion rate of the Universe.  Measurements of $H_0$ from different
astronomical observations are at odds with each other. For instance, $H_0$
inferred from the fluctuation spectrum of the cosmic microwave
background~\citep{Aghanim:2018eyx} disagrees with the value obtained from the
measurement of the luminosity distance and redshift to Type Ia
supernovae~\citep{Riess:2019cxk, Freedman:2019jwv, Wong:2019kwg} at
$4.0$--$5.8\sigma$ significance~\citep{Riess:2019cxk, Verde:2019ivm}.
Confirming or ruling out this discrepancy is of paramount importance as it may
point to new or missing physics from an epoch in the early Universe just before
the recombination era \citep{Verde:2019ivm}. 

Gravitational waves (GWs) facilitate a unique way of determining $H_0,$ without
relying on the cosmic distance ladder~\citep{Schutz86, Holz:2005df}. The
multi-messenger observations of the binary neutron star merger
GW170817~\citep{GW170817} led to the first GW-assisted measurement of $H_0$,
estimating its value to be $70^{+12}_{-8} \,{\rm km}\,{\rm s}^{-1}\,{\rm
Mpc}^{-1}$~\citep{GW170817_H0} (also see
\cite{Hotokezaka:2018dfi,Dhawan:2019phb}). This measurement crucially relied on
the coincident detection of an electromagnetic (EM) counterpart to the GW
source. The source's redshift came from the counterpart while the luminosity
distance was inferred from the GW signal. Observation of EM counterparts to
$\sim 50$ binary neutron star mergers could nail down the \hubble constant to
an accuracy of 2\%, sufficient to confirm if there are any systematics in the
local measurements of $H_0$ \citep{Chen:2017rfc} (also see
\cite{Feeney:2018mkj,Mortlock:2018azx}).

\section{$H_0$ with Dark Sirens}
Binary black hole (BBH) mergers are not expected to have EM counterparts but
they too can determine the luminosity distance to their hosts independently of
the cosmic distance ladder; for this reason they are sometimes referred to as
\emph{dark sirens}. Even so, it may be possible to identify their potential
host galaxies either with the help of a galaxy catalog or by follow-up
observations \citep{Nishizawa:2016ood,Yu:2020vyy}. There is no guarantee that
this approach could identify the true host as the GW sky localization with
current generation detectors is not precise enough \citep{Aasi:2013wya}.  With
multiple potential hosts one has to resort to a statistical approach to
determine $H_0$ as first suggested in \cite{Schutz86} (also see
\cite{DelPozzo:2011yh}). Dark sirens detected in the first and second observing
runs of LIGO and Virgo \citep{LIGOScientific:2018mvr} were used in this way to
estimate the value of $H_0$ to be $68^{+14}_{-7}\,\rm km\, s^{-1}\, Mpc^{-1}$
\citep{Abbott:2019yzh} (also see \cite{Soares-Santos:2019irc,Palmese:2020aof}).  

One source of systematic errors in this method arises from the incompleteness
of the available galaxy catalogs. Three galaxy catalogs were used in
\citep{Abbott:2019yzh}: The first one is the GLADE catalog \citep{Dalya:2018cnd},
which has an all-sky coverage but the probability of the host galaxy to be in
the catalog at $z=0.1$, as determined in \cite{Abbott:2019yzh}, is $\sim 60\%$.
The other two catalogs are from the Dark Energy Survey (an ongoing five year
survey) \citep{Drlica-Wagner:2017tkk,Abbott:2018jhe} and GWENS \citep{GWNES_ref},
which are mostly complete up to $z=0.1,$ but do not cover the complete sky,
with the former covering only an eighth of the sky at the end of its mission.
Nonetheless, LIGO and Virgo at their design sensitivity could constrain $H_0$
with dark sirens to 5\% accuracy with $\sim250$ detections~\citep{Gray:2019ksv}
(also see \cite{Nair:2018ign}).  

\section{Motivation for Current Work}

Recently, LIGO and Virgo have published two compact binary mergers found during
the third observing run: GW190412~\citep{LIGOScientific:2020stg} and
\gwOhEightS~\citep{Abbott:2020khf}, which are exceptional due to their large
mass-asymmetry, with mass ratios $\sim 3$ and $\sim 9$, respectively. These
systems have led to the detection of subdominant spherical harmonic modes
beyond the quadrupole mode
\citep{Roy:2019phx,LIGOScientific:2020stg,Abbott:2020khf}. Such higher modes
have been argued to be of importance in the parameter estimation of asymmetric
binaries \citep{ChrisAnand06b, AISSV07, Graff:2015bba}, especially that of the
luminosity distance $D_L$ and orbital inclination $\iota$~\citep{Ajith:2009fz}.
Indeed, among all the dark compact binaries detected till date, \gwOhEightS has
the best measured luminosity distance ($\sim 18\,\%$) and 90\% credible sky
area ($\sim 19\,\text{deg}^2$)  \citep{Abbott:2020khf}, albeit its signal to
noise ratio (SNR) is similar to that of GW150914 whose uncertainty in distance
is $\sim 35\%$~\citep{LIGOScientific:2018mvr}. The first LIGO-Virgo GW
transient catalog \gwtc~\citep{LIGOScientific:2018mvr} includes ten BBHs, of
which some are broadly similar to \gwOhNine. Future observations could
potentially improve parameter estimations of such systems when higher modes are
present in the observed signal. 

We show that the recent discoveries have raised the opportunity to measure
$H_0$ with dark sirens to within 2\%, the accuracy required to resolve the
Hubble tension, in the next five years. There are two principal reasons for
this expectation: (1) planned upgrades to LIGO and Virgo would enhance their
sensitive volume by a factor of $\sim 3.4$, and (2) higher spherical harmonic
GW modes, as shown in the Appendix, should help localize dark sirens in the sky
by a factor of $\sim 2$ better but, more critically, reduce the uncertainty in
distance measurement \citep{Ajith:2009fz, Harry:2017weg, Kalaghatgi:2019log} by
a factor as large as 6 for the populations we study below. These improvements
compound together to localize a `golden' subset of the dark sirens to a small
enough patch in the sky that only a single galaxy would be found within the
error region. Galaxy catalogs and EM follow-up campaigns could then determine
the host, obtain the source's redshift, and hence directly measure $H_0,$
without relying on the statistical method \citep{Howell:2017wvf,Kuns:2019upi}. 

We are primarily interested in the local measurement of $H_0$ with sources
close enough that we can neglect the effect of weak lensing \citep{Holz:2005df}
as well as dark matter and dark energy and assume the simplest form of the
\hubble law. If sources are too close (say, $D_L\lesssim 100\,\rm Mpc)$, $H_0$
measurements will be flawed due to the systematic bias from peculiar velocities
$v_p$ of host galaxies. While galaxies in clusters have relatively large
peculiar velocities ($v_p \sim 2000\,\rm km\,s^{-1}$), this appears not to be
the case ($v_p \sim 300\,\rm km\,s^{-1}$) for the majority of galaxies
(${\sim}95\%$) that are found outside of clusters \citep{Bahcall:1995tf}.
Therefore, we will consider, in our study, dark sirens distributed uniformly in
co-moving volume up to a redshift of $z=0.1$, corresponding to a luminosity
distance $D_L\simeq 475\,\rm Mpc$ for the \emph{Planck 2015}  cosmology
\citep{Ade:2015xua}. This ascertains that most sources are far enough away
($\sim1\%$ will be closer than $D_L\sim100\,\rm Mpc$) that the \hubble flow
will dominate the peculiar velocities.

\begin{table}[t]
\centering
    \begin{tabular}{p{0.25\columnwidth}|p{0.70\columnwidth}}
    \hline \hline 
    Network label & Detector location (technology) \\
    \hline
    \hline
    \plusNetThree & Hanford WA (A+), Livingston LA (A+), Cascina Italy (AdV+) \\
    \hline
    \plusNet & Hanford WA (A+), Livingston LA (A+), Cascina Italy (AdV+), Kamioka Japan (KAGRA+), Hingoli India (A+) \\
    \hline
    \voyNet & Hanford WA (Voyager), Livingston (Voyager), Cascina Italy (AdV+), Kamioka Japan (KAGRA+), Hingoli India (Voyager) \\
    \hline
    \initNet & Cascina Italy (ET-D), fiducial US site (CE1), fiducial Australian site (CE1) \\
    \hline \hline
    \end{tabular}
    \caption{An overview of the four networks used to benchmark GW detections in the study. The location determines the detector antenna patterns and location phase factors, while the technology indicates the used power spectral density. The Voyager and Cosmic Explorer power spectral densities are chosen to be low-frequency optimized and in the case of the latter for a detector arm length of $40\,\text{km}$.}
    \label{tab:networks}
\end{table}

\section{Dark Siren Populations for $H_0$} 

We consider three types of dark siren populations in our analysis and compute
the precision with which $H_0$ could be measured with golden binaries among
those populations. The first population makes use of the rates and mass
distributions inferred from the \gwtcS BBHs \citep{LIGOScientific:2018mvr,
LIGOScientific:2018jsj}.  The second population represents the \emph{heavy BBH}
sub-population of \gwtcS with companion masses larger than $25 M_\odot$. The
binaries in this population might be among the loudest that GW detectors can
observe. Their projected merger rate makes them an interesting target to
consider in our study.  Lastly, we consider a population of
\emph{\gwOhEight-like} dark sirens. The interest in this class of sources stems
from the expected role higher modes may play in parameter estimation. 

The simulated populations differ in the choice of companion masses, which are
all specified in the source frame in this study. In the case of the full \gwtcS
BBH population, we distribute the larger mass $m_1\in[5 M_{\odot},100
M_{\odot}]$ according to a power-law $p(m_1)\propto m_1^{-\alpha}$ with
exponent $\alpha=1.6$ and the smaller mass $m_2$ uniformly in the range $[5
M_{\odot},m_1]$ \citep{LIGOScientific:2018jsj}. The heavy BBH population only
differs by increased lower-mass bounds: $m_1,m_2 \geq 25 M_\odot$. Finally, the
component masses are fixed to $m_1=23\,M_{\odot}$ and $m_2=2.6\,M_{\odot}$ for
\gwOhEight-like events \citep{Abbott:2020khf}.  In all the cases, the companion
black holes are assumed to be non-spinning, consistent with \gwtc.  Further,
the events are uniformly distributed in co-moving volume, according to
\citep{Ade:2015xua}, up to redshift $z=0.1$, as well as over sky positions, and
orientation angles.  Each simulated population contains $10^4$ samples.

We consistently employ the \texttt{IMRPhenomHM}~\citep{London:2017bcn} model
from \texttt{lalsimulation}~\citep{lalsuite} for the parameter estimation of all
the aforementioned classes of sources. This waveform family includes radiative
moments with spherical harmonic indices $(\ell, m)= (2, 2), (3, 3), (4, 4), (2,
1), (3, 2), (4, 3),$ which ensures that there are no systematic biases due to
the neglect of higher modes and helps us carryout meaningful comparisons
between the three populations. 

The median merger rate for each population reported by LIGO and Virgo is
$R_\text{\gwOhEight-like}=7^{+16}_{-6}\,{\rm Gpc}^{-3}\, {\rm yr}^{-1}$ for
\gwOhEight-like events \citep{Abbott:2020khf} and
$R_\text{\gwtc}=53^{+59}_{-20}\,{\rm Gpc}^{-3}\, {\rm yr}^{-1}$ for \gwtcS BBHs
\citep{LIGOScientific:2018jsj}.  Since we restrict the populations to a maximum
luminosity distance of $D_L=475\,\mathrm{Mpc}$, we obtain the median merger
rates in the volume of interest for these populations to be
$\bar{R}_\text{\gwOhEight-like}=3.1\,{\rm yr}^{-1}$ and
$\bar{R}_\text{\gwtc}=24\,{\rm yr}^{-1}$. Consequently, we obtain
$\bar{R}_\text{heavy}=2.9\,{\rm yr}^{-1}$ for heavy BBH events, where we assume
that $\bar{R}_\text{heavy}= f\,\bar{R}_\text{\gwtc}$ and $f$ to be the fraction
of \gwtcS BBHs with $m_1,m_2\geq25\,M_\odot$: $$ f \simeq 1.9
\int_{25}^{100}\mathrm{d}m_1\,m_1^{-1.6}\int_{25}^{m1}\mathrm{d}m_2\,\frac{1}{m_1-5}
\simeq 0.12.  $$ These rates will be used below to estimate the number of dark
sirens that could be localized well enough each year to identify their hosts.

\begin{figure*}
    \centering
    \includegraphics[width=\linewidth]{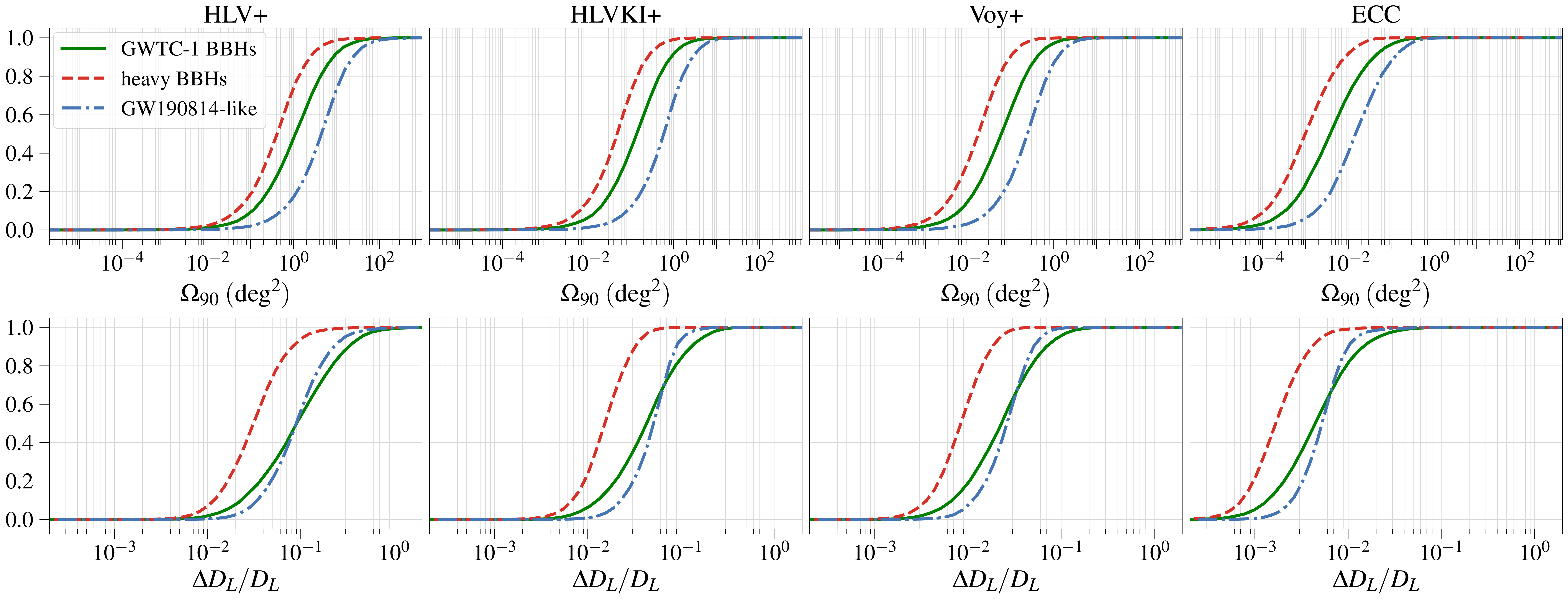}
    \caption{Cumulative density functions of the 90\% credible sky area
    $\Omega_{90}$ and fractional luminosity distance error $\Delta D_L / D_L$
    for three dark siren populations (\gwtcS BBHs, heavy BBHs, and
    \gwOhEight-like) in the four studied networks. The populations contain
    $10^4$ binaries. The crossover in the curves of the \gwtcS BBH and
    \gwOhEight-like populations in the lower panels is likely due to the
    trade-off between measurement accuracies obtained from massive binaries and
    asymmetric ones.}
    \label{fig:distance_sky_histogram}
\end{figure*}

\section{Detector Networks}

The detector networks considered in this study are combinations of seven
geographical locations and three technology generations (essentially, the
choice of detector's power spectral density) as summarized in
Table~\ref{tab:networks}.  A+, AdV+, and KAGRA+ \citep{aPlusWP,AdV_KAGRA_plus}
are planned upgrades of the Advanced LIGO \citep{aLIGO_ref}, Advanced Virgo
\citep{AdV_ref}, and KAGRA \citep{Akutsu:2018axf} detectors, referred to as
2G+.  Their targeted strain sensitivity should improve the reach for BBHs by a
factor of $\sim1.5$. We consider two 2G+ networks, \plusNetThreeS and \plusNet,
with three and five detectors, respectively.  The third network, \voyNet, is
heterogeneous and uses two 2G+ and three `2.5G' detectors; the latter is a
proposed upgrade of the LIGO facilities to `Voyager' technology
\citep{voyager_ref}, which will introduce a further improvement in their reach
by a factor of $\sim3.2$ compared to 2G+.  The final network, \initNet,
contains three third generation (3G) observatories, namely one Einstein
Telescope \citep{ET_ref} in Italy and two Cosmic Explorer detectors
\citep{2019arXiv190704833R} at fiducial sites in the United States and
Australia. The used power spectral densities, ET-D and CE1
\citep{AdV_KAGRA_plus}, yield strain sensitivity improvements by factors of
$\sim3$ and $\sim5$, respectively, as compared to Voyager.

\section{Measurement Accuracies}

For each of the above networks, we compute numerically the error in the
estimation of the binary parameters using the Fisher-matrix formalism
\citep{CF94,PW95}, which is an excellent approximation for the high-SNR events
such as the ones in our three populations.  The parameter set includes the
binary's intrinsic masses, $m_1$ and $m_2$, two angles describing the position
of the binary in the sky (right ascension $\alpha$ and declination $\delta$),
two more giving its orientation relative to the detector (inclination  $\iota$
of the binary's orbital angular momentum relative to the line of sight and the
polarization angle $\psi$), the luminosity distance $D_L$, a fiducial
coalescence time, and the phase of the signal at that time. Among these, $D_L$
and $\iota$ are highly correlated when the observed signal is face-on and
contains only the $\ell=2$ quadrupole mode but the degeneracy is largely lifted
by higher order spherical harmonic modes, $\ell > 2$ \citep{Ajith:2009fz,
CalderonBustillo:2020kcg}. Gravitational waves from coalescing binaries are
dominated by the quadrupole mode, however, higher modes are present in systems
with unequal mass companions and more prominent for systems observed with large
inclination angles. 

Figure~\ref{fig:distance_sky_histogram} shows the cumulative density functions
of the 90\% credible sky area $\Omega_{90}$ and fractional luminosity distance
error $\Delta D_L / D_L$ for each of the three dark siren populations in the
four studied networks. The heavy BBH population allows for better constrains on
$D_L$ and $\Omega_{90}$ due to the large SNRs such massive mergers would
accumulate. The \gwOhEight-like events will be able to determine the luminosity
distance to similar accuracies as \gwtcS BBHs, but fall behind in terms of the
sky localization.  The improved parameter estimation, for \gwOhEight-like
signals with relatively low SNRs, is facilitated by the higher modes that are
strongly excited for such highly asymmetric systems.

\begin{table}
\centering
    \begin{tabular}{c|c|c|c|c}
    \hline \hline 
    \multicolumn{1}{c|}{Metric} & \plusNetThree & \plusNet & \voyNet & \initNet  \\
    \hline
    \hline
    \multicolumn{5}{c}{\multirow{2}{*}{\emph{\gwtcS BBH population} --- $\bar{R}_\text{\gwtc}=24\,{\rm yr}^{-1}$}} \\ 
    \multicolumn{5}{c}{} \\
    \hline
    \hline
    $\langle \rho \rangle$ & 160 & 150 & 260 & 760  \\
    \hline
    $\langle \Delta D_L / D_L \rangle $~~$(\%)$ & 3.2 & 2.3 & 1.5 & 0.43  \\
    \hline
    \hline
    $\epsilon^*$~~(\%) & 4.3 & 24 & 43 & 91  \\
    \hline
    Event rate~~$(\text{yr}^{-1})$ & 1.0 & 5.8 & 10 & 22  \\
    \hline
    \hline
    \multicolumn{5}{c}{\multirow{2}{*}{\emph{heavy BBH population} --- $\bar{R}_\text{heavy}=2.9\,{\rm yr}^{-1}$ }} \\ 
    \multicolumn{5}{c}{} \\
    \hline
    \hline
    $\langle \rho \rangle$ & 250 & 250 & 460 & 1900  \\
    \hline
    $\langle \Delta D_L / D_L \rangle $~~$(\%)$ & 1.9 & 1.3 & 0.79 & 0.17  \\
    \hline
    \hline
    $\epsilon^*$~~(\%) & 9.6 & 48 & 76 & 100  \\
    \hline
    Event rate~~$(\text{yr}^{-1})$ & 0.28 & 1.4 & 2.2 & 2.9  \\
    \hline
    \hline
    \multicolumn{5}{c}{\multirow{2}{*}{\emph{\gwOhEight-like population} --- $\bar{R}_\text{\gwOhEight-like}=3.1\,{\rm yr}^{-1}$}} \\ 
    \multicolumn{5}{c}{} \\
    \hline
    \hline
    $\langle \rho \rangle$ & 63 & 75 & 160 & 530  \\
    \hline
    $\langle \Delta D_L / D_L \rangle $~~$(\%)$ & 4.3 & 2.7 & 1.5 & 0.51  \\
    \hline
    \hline
    $\epsilon^*$~~(\%) & 1.1 & 5.3 & 14 & 75  \\
    \hline
    Event rate~~$(\text{yr}^{-1})$ & 0.034 & 0.16 & 0.43 & 2.3  \\
    \hline
    \hline
    \end{tabular}
        \caption{The \emph{medians} of the signal to noise ratio $\langle\rho\rangle$ and fractional error in luminosity distance $\langle\Delta D_L / D_L\rangle$ for three dark siren populations, \gwtcS BBHs, heavy BBHs, and \gwOhEight-like. The medians are computed for the fraction $\epsilon^*$ of events in our simulation that satisfy the condition $\Omega_{90} \lesssim \Omega^* = 4.4\,\times \, 10^{-2} \,\text{deg}^2$, in the respective network. The Table also lists the fraction $\epsilon^*$ and the corresponding number of events per year, for sources within a redshift of $z\leq0.1$. The addition of two detectors in \plusNetS compared to \plusNetThreeS improves the sky localization for all events, thus allowing quieter signals to fulfill $\Omega_{90} \lesssim \Omega^*$ and thus resulting in a decreased median SNR for the \gwtcS population, see Figure \ref{fig:hm_vs_d_all} in the Appendix.}
    \label{tab:populations_bench}
\end{table}

These findings present the two quantities in an ``event-independent'' fashion.
We tackle this by applying a sky localization condition to each event. A sky
patch of size $\Omega^*\simeq 4.4\times10^{-2}\,\text{deg}^2$ contains, on
average, one galaxy with luminosity $L\geq L_{10}=10^{10}\, L_\odot$ within
$z=0.1$ (refer to Eq. (7) in \cite{Singer:2016eax}). Thus, from the full set of
simulated events, we select the fraction $\epsilon^*$ that is resolved to
$\Omega_{90} \lesssim \Omega^*$. This ensures a unique identification of the
dark siren's host galaxy. Table~\ref{tab:populations_bench} lists, for this
sub-population of events, \emph{medians} of the SNR $\langle \rho \rangle$ and
the fractional error in the luminosity distance $\langle \Delta D_L /
D_L\rangle$. Furthermore, from the merger rate $\bar R$ and the fraction
$\epsilon^*$, we compute the number of well-localized dark sirens detected each
year by the different networks.  The listed event rates suggest that we can
expect to observe one to several such events from the \gwtcS and heavy BBH
populations every \emph{two years}. \gwOhEight-like binaries will be rarer at
only one event every 25, 6.3, and 2.3 years in \plusNetThree, \plusNet, and
\voyNet, respectively.  The golden binaries described in
Table~\ref{tab:populations_bench} should all yield an error $\leq5\%$ in the
luminosity distance, with the most accurate distance estimates to be expected
from heavy binaries. Finally, the values in Table \ref{tab:populations_bench}
clearly show that dark siren localization, both in terms of luminosity distance
measurement and host galaxy identification, will be the norm in the 3G network
era (\initNet). In fact, the distance will be determined to sub-percent
accuracy no matter the source population.

\section{Measurement of $H_0$ with dark sirens}

\begin{figure*}
    \centering
    \includegraphics[width=2\columnwidth]{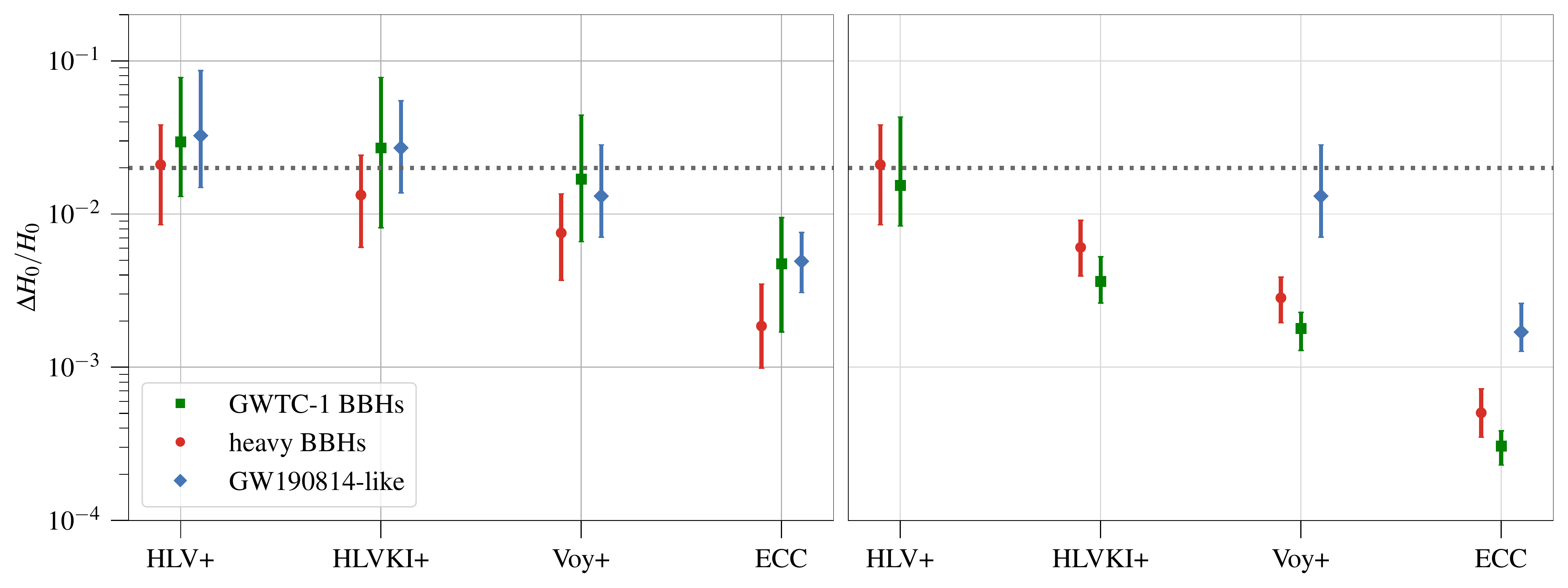}
    \caption{\emph{Left panel:} The median and standard deviation of the distribution of fractional errors in $H_0$ measurement (assuming errors from redshift measurements are negligible) over 100 realizations for a single dark siren event for the four networks under study. All the three populations we considered give statistically similar estimates of $H_0$. 
    \emph{Right panel:} Same as the left panel but the errors have been computed for the expected number of events in two years of observing time for a network, assuming 100\% duty cycle. The error estimates for some of the cases are not plotted because we do not expect to see any such event in an observing time of two years.}
    \label{fig:h0} 
 \end{figure*}

The luminosity distance--redshift relation in the local Universe is well
approximated by the relation $D_L = cz/H_0.$ It follows, then, that the
fractional error in $H_0$ is equal to the fractional error in $D_L$ for a
single event, if errors in redshift measurements are negligible. One source of
systematic error that affects the determination of $H_0$ is the selection bias
that enters the analysis due to the consideration of only a sub-population of
actual events based on the events' sky localization. For the distances
considered in this study, the selection bias re-weights the $H_0$ posterior by
a factor of $1/H_0^3$ \citep{Chen:2017rfc}. For the expected value of $H_0$ and
the errors of interest in this study, the effect of this bias is to move the
$H_0$ peak to lower values and widen the width of the posterior at the level of
a percent for the worst errors quoted. In a Fisher study, only the errors in
the estimation of a parameter are important and not the position of the peak of
the posterior distribution. Consequently, we neglect the effects of the
selections bias. In the left panel of Figure~\ref{fig:h0}, we show the error in
the measurement of $H_0$ by a single, golden event for each of the three dark
siren sub-populations that obeys the localization condition $\Omega_{90} <
\Omega^*.$ We plot the median and variance of a distribution of $H_0$ errors
for 100 realizations of a single random event drawn from each sub-population.
We find that HLV+, HLVKI+, and Voy+ detectors would estimate $H_0$ to a few
percent accuracy while the 3G network would measure $H_0$ to sub-percent
precision.

A major factor that contributes to the $D_L$ errors is the $D_L$-$\iota$
degeneracy \citep{Ajith:2009fz}. This is most significant in the absence of
strongly excited higher modes and, therefore, affects the nearly equal-mass
binaries in the \gwtcS and heavy BBH populations the most. It is precisely due
to the importance of higher modes that golden \gwOhEight-like events stay
competitive to much higher-SNR signals from massive binaries. This holds
especially true for less sensitive networks which are less effective at
overcoming the $D_L$-$\iota$ degeneracy due to lower SNRs. In fact, the
variances in the distribution of $H_0$ errors do not favor any of the three
dark siren populations in the 2G+ networks. It is not until the 3G era, when
the heavy BBH events with high-SNRs will yield significantly tighter bounds on
$H_0$ than the \gwOhEight-like ones.

Further, we note that the variance is larger for the \gwtcS population compared
to the heavy BBH and \gwOhEight-like populations. This can be attributed to the
wide mass distribution of the \gwtcS population which ranges from $5M_{\odot}$
to  $100M_{\odot}$: lower mass binaries with relatively low SNR result in
poorer luminosity distance measurements, but could still fulfill the sky
localization condition. However, the \gwtcS dark sirens stay competitive to the
other two populations since they include both massive, high-SNR systems as well
as very asymmetric ones.

In the right panel of Figure~\ref{fig:h0}, we calculate the errors in $H_0$ by
taking account of the number of detections (rounding to the nearest integer) in
two years of observing time for each network (assuming 100\% duty cycle). The
$H_0$ error estimates are missing for \gwOhEight-like dark sirens in the case
of \plusNetThree and \plusNet, since we do not \emph{expect} to observe
\emph{well-localized} dark sirens of either type in the respective networks
within a two year time period. We see that, due to the different rates of
detections for different populations, the performance of the \gwOhEight-like
population is slightly worse (about a factor of 2) than the other two
populations. Note also that the heavy BBH population is still competitive with
the GWTC-1 BBHs even though its rate is considerably lower than the latter.

\section{Conclusions}

We have demonstrated a tantalizing possibility of measuring the \hubble
constant to $\sim$2\%-level precision using dark sirens with the imminent
upgrades of the LIGO and Virgo detectors to 2G+ sensitivity.  For highly
asymmetric, low-mass systems like \gwOhEight, the inclusion of higher spherical
harmonic modes is crucial to make such a measurement especially in the A+ era,
while heavier systems see less stark improvements, see Appendix.   

Our conclusions rely on controlling the amplitude calibration of the detectors
to  below $\lesssim 1\%$, which can be accomplished with photon calibrators
\citep{Karki:2016pht}, and  two assumptions on a dark siren's host galaxy: its
unique identification and a negligible uncertainty in peculiar velocity
correction. The sky area could be `contaminated' with faint galaxies or the
host itself could be faint and missing from current catalogs. Further, the
host's peculiar velocity correction might not meet the desired accuracy,
especially for close-by sources with small \hubble flow. Fortunately, EM
follow-up observations of such well-localized sirens should be able to identify
the host galaxy, obtain the redshift accurately by spectroscopy, model the
velocity flow, and constrain the uncertainty in peculiar velocities to $\sim
100$ -- $150 \, \text{km/s}$ \citep{Mukherjee:2019qmm}, which is accurate enough
for ~99\% of the sources considered in this study. Such follow-up surveys will
be of interest to the entire astrophysics community since they would not only
benefit from the \hubble constant measurement, but improve our understanding of
the correlations between binary coalescences and their environments. 

Given the paucity of binary neutron star mergers with EM counterparts so far,
dark sirens offer an alternative to resolve the $H_0$ tension within the next
five years \citep{aPlusWP}. Beyond the 2G+ era, our results are also very
encouraging for a possible synergy between the dark sirens and the bright
sirens, wherein the $H_0$ measurement from low redshift may be used as a prior
in the measurement of other cosmological parameters at higher
redshifts~\citep{Sathyaprakash:2009xt}.

\section{Acknowledgements} 

We thank Archisman Ghosh, Leo Singer, and Salvatore Vitale for useful
discussions and E. Hall and K. Kuns for providing Cosmic Explorer sensitivity
curves. We thank Martin Hendry for carefully reading the manuscript and
providing useful comments.  B.S.S. is supported in part by NSF Grant No.
PHY-1836779, AST-1716394 and AST-1708146. S.B. and A.D. are supported by NSF
Grant No. PHY-1836779. K.G.A. is partially supported by a grant from the
Infosys Foundation. K.G.A. acknowledges the Swarnajayanti grant
DST/SJF/PSA-01/2017-18 DST-India and Core Research Grant EMR/2016/005594 of
SERB. We thank all front line workers combating the CoVID-19 pandemic without
whose support this work would not have been possible. This paper has the LIGO
document number LIGO-P2000229.

\clearpage

\appendix

\section{Impact of higher mode waveforms on detectability, sky localization, and distance measurement}

\begin{figure*}[hb]
\includegraphics[width=\textwidth]{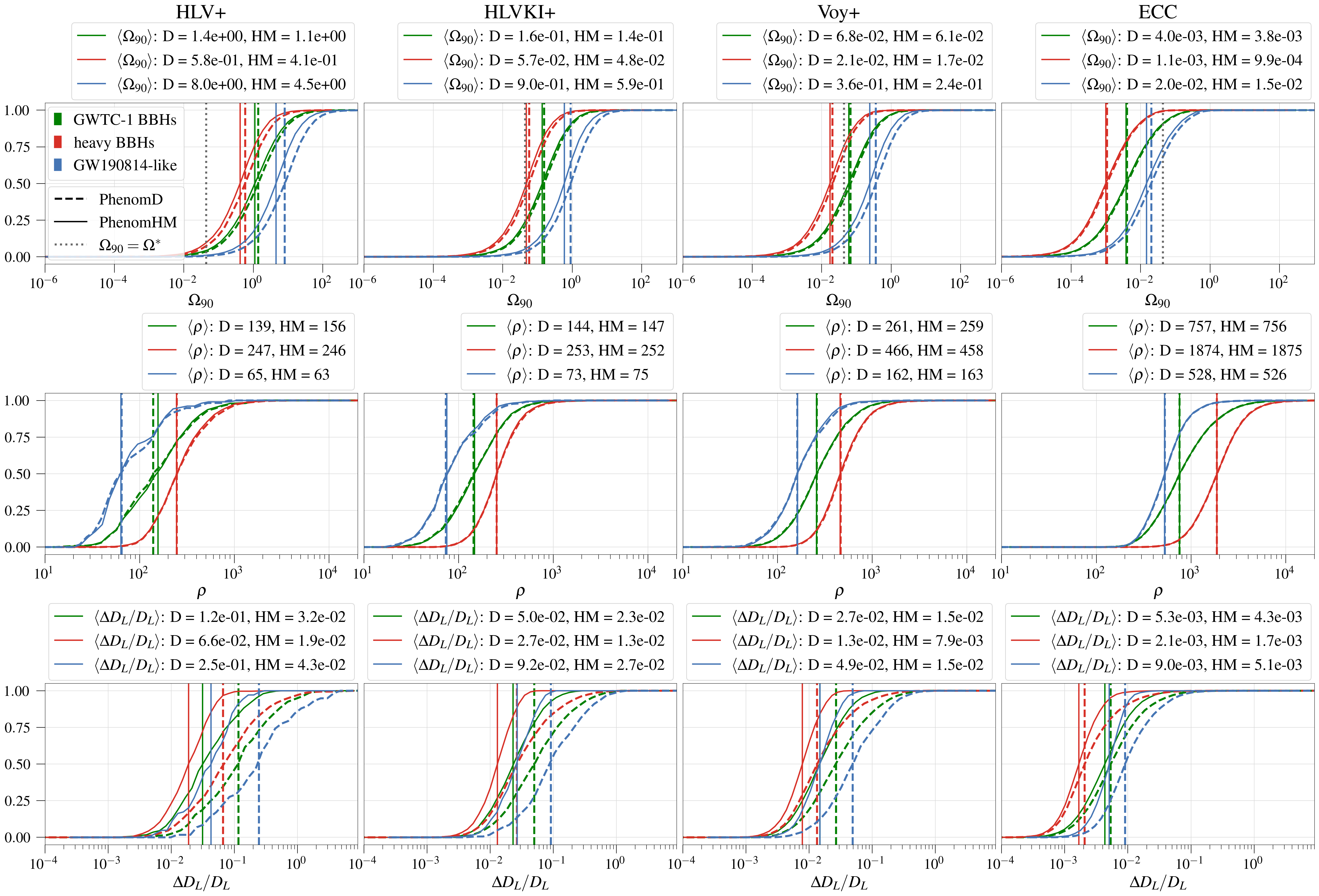} \\
\caption{Cumulative density functions and medians of the 90\% credible sky area $\Omega_{90}$, SNR $\rho$, and
fractional luminosity distance error $\Delta D_L /D_L$ calculated with two waveform models, one
including higher modes (PhenomHM, solid) and the other without (PhenomD,
dashed), for the \gwtcS BBH, heavy BBH, and
\gwOhEight-like populations. SNRs and distance errors are restricted to systems fulfilling
$\Omega_{90}<\Omega^*$.}
\label{fig:hm_vs_d_all}
\end{figure*}

Figure \ref{fig:hm_vs_d_all} presents the cumulative density functions and
medians of the 90\% credible sky area $\Omega_{90}$, SNR $\rho$, and fractional
error in luminosity distance $\Delta D_L/D_L$ for the \gwtcS BBH, heavy BBH,
and \gwOhEight-like populations, estimated with two different waveform models:
one with higher modes, PhenomHM \citep{London:2017bcn}, and one without, PhenomD
\citep{Husa:2015iqa}. We restricted the SNR and distance errors to the binaries
in each population that fulfill the sky localization condition
$\Omega_{90}<\Omega^*=4.4\times10^{-2}\,\mathrm{deg^2}$ for either waveform.
The graphs and quoted median values clearly illustrate the impact that
waveforms with higher modes have on the systematics of a measurement: SNRs are
not affected in a meaningful way, but both sky localization and distance
measurements show improvements. This holds especially true in the A+ era. The
median 90\%-credible sky areas shrink by $\sim$20\%--45\% with higher mode
contributions and the average distance errors improve by factors of $\sim$3--6.
Even in a 3G network, we expect to observe an enhancement of the order
$\sim$5\%--25\% in sky localization and $\sim$20\%--45\% for the distance
measurement when including higher modes. Hence, these figures demonstrate the
importance of the contribution of higher modes to achieve an accurate
measurement of the sky location and luminosity distance, and consequently the
Hubble constant.

\bibliographystyle{yahapj}
\bibliography{cosmic-rung}
\end{document}